\documentclass[aps,floats,twocolumn,showpacs]{revtex4-1}
\usepackage{amssymb}
\usepackage{amsmath}
\usepackage{epstopdf}
\usepackage{graphicx}

 \def\bc{\begin{center}}          \def\ec{\end{center}}
 \allowdisplaybreaks
\begin{document}
 \title{Excitation of two-dimensional plasma wakefields by trains of equidistant particle bunches}
 \author{K.V.Lotov}
 \affiliation{Budker Institute of Nuclear Physics SB RAS, 630090, Novosibirsk, Russia}
 \affiliation{Novosibirsk State University, 630090, Novosibirsk, Russia}
 \date{\today}
 \begin{abstract}
Nonlinear effects responsible for elongation of the plasma wave period are numerically studied with the emphasis on two-dimensionality of the wave. The limitation on the wakefield amplitude imposed by detuning of the wave and the driver is found.
 \end{abstract}
 \maketitle

Plasma wakefield acceleration (PWFA) driven by charged particle beams is one of many advanced accelerating techniques aimed at future ultimate energy or compact particle accelerators. Originally, it was proposed to excite the plasma wave for PWFA by a train of equally spaced short electron bunches \cite{Chen}. Later, the focus has shifted to single electron bunches operating in the so-called blowout regime \cite{PRA44-6189,PoP9-1845,Nat.445-741}.

Recently, the interest in multi-bunch wave excitation was regenerated in the context of the proton-driven PWFA. The existing proton beams have huge energy contents. It is attractive to use these beams as drivers for boosting electrons well beyond nowadays energy frontiers \cite{NatPhys9-363,PoP18-103101}. However, for efficient excitation of the wakefield, the proton bunch must be reshaped to have a longitudinal structure of the plasma wavelength scale \cite{PRST-AB13-041301}. Reshaping by conventional methods is too expensive and inefficient, so it was proposed to use the plasma for chopping the initially long proton bunch into a train of short micro-bunches \cite{PPCF53-014003}. The mechanism of chopping is the self-modulation instability that produces the micro-bunches spaced one plasma period apart \cite{PRL104-255003}. The question thus arises of how strong wakefield can be achieved with trains of equidistant particle bunches. In the paper we address this question for beam parameters of interest for the proton-driven PWFA.

A similar question arose in the context of plasma beatwave acceleration (PBWA) \cite{Taj}. In PBWA, the plasma wave is produced by the beating of two laser beams whose frequencies differ by approximately the plasma frequency $\omega_p = \sqrt{4 \pi n_0 e^2/m}$, where $n_0$ is the plasma density, $e$ is the elementary charge, and $m$ is the electron mass. The wave growth saturates because of a nonlinear shift in the plasma wave frequency \cite{PRL29-701,PF30-904,RMP81-1229}. We show that the limitation observed in PWFA is qualitatively the same, but quantitatively is different for several reasons.

\begin{figure}[b]
\bc\includegraphics[width=186bp]{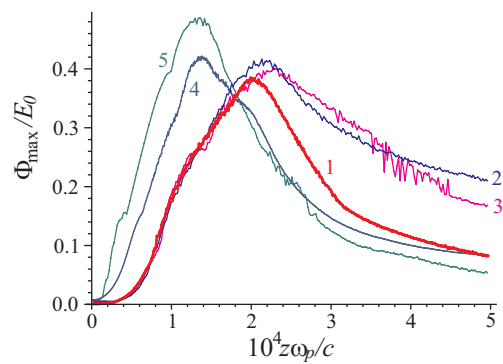} \ec
 \vspace*{-5mm}
\caption{(Color online) Dimensionless wakefield amplitude versus the dimensionless propagation distance for various self-modulating proton beams.}\label{f1-growth}
\end{figure}

\begin{figure}[t]
\bc\includegraphics[width=225bp]{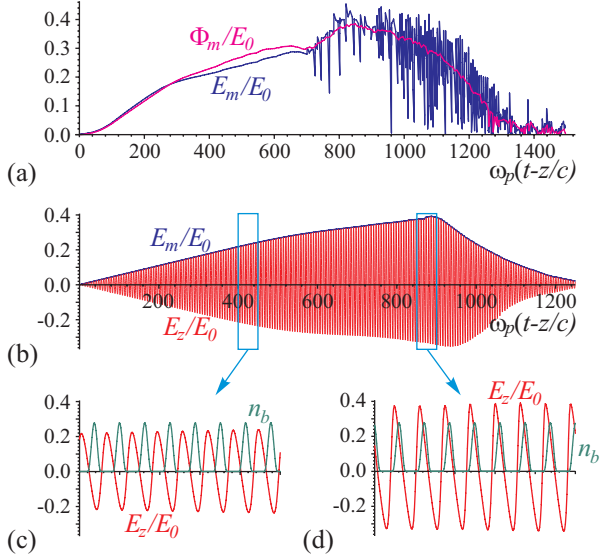} \ec
 \vspace*{-5mm}
\caption{(Color online) Temporal growth of the wakefield (envelopes $\Phi_m$ and $E_m$ of the wakefield potential $\Phi$ and electric field $E_z$) for the self-modulated AWAKE beam at $z=4\,$m (a) and for the test train of rigid bunches of the same peak density (b-d); zoomed fragments (c) and (d) also show the beam density $n_b$ in arbitrary units. }\label{f2-envelopes}
\end{figure}

This study was motivated by the observation that the wakefield growth in simulations of self-modulating proton beams saturates at about 40\% of the wavebreaking field $E_0 = m c \omega_p/e$, where $c$ is the speed of light. Substantial variations of beam parameters do not result in corresponding increase of the peak field. To illustrate this statement, we show in Fig.\,\ref{f1-growth} the wakefield excited by various self-modulating beams in the plasma. As a measure of the wakefield amplitude, we take the maximum value $\Phi_\text{max} (z)$ of the wakefield potential $\Phi (z,t)$ on the beam axis:
\begin{equation}\label{e1}
    \Phi (z,t) = \omega_p \int_{-\infty}^t E_z(z, t') \, d t',
\end{equation}
where $E_z$ is the on-axis electric field. It is more convenient to characterize the wakefield by the potential rather then by the electric field itself since the latter has a singular character for strongly nonlinear waves, and the observed envelope of $E_z$ is noisy [Fig.\,\ref{f2-envelopes}(a)] and depends on the resolution of simulation codes and on the plasma temperature \cite{PRST-AB6-061301}. Simulations are made with the axisymmetric kinetic version of the quasi-static code LCODE \cite{PRST-AB6-061301,IPAC13-1238}

The curve `1' in Fig.\,\ref{f1-growth} corresponds to the baseline variant of AWAKE experiment at CERN \cite{CDR,IPAC13-1179}. The main parameters of the proton beam are: energy 400\,GeV, radius $\sigma_z = 0.2$\,mm, length $\sigma_z = 12$\,cm, peak density $n_{bm} = 4\times 10^{12}\,\text{cm}^{-3} = 0.0057\,n_0$, normalized emittance 3.6\,mm\,mrad. The plasma density is $n_0=7 \times 10^{14}\,\text{cm}^{-3}$, so that $c/\omega_p=\sigma_r$. The beam is half-cut at the midplane for seeding the instability.

The curve `2' is plotted for the opposite charge beam (antiprotons) with the same parameters. Usually, negatively charged drivers excite the wave more efficiently \cite{AIP396-75,PRE64-045501}. This asymmetry, indeed, shows up at the stage of field decay, but not at the field maximum. The case `3' differs from the baseline case `1' by 10 times smaller beam emittance. A low emittance has a favorable effect on the wakefield in general, but not at the field maximum. Curve `4' corresponds to the longitudinally compressed beam with 4 times higher peak density. However, the maximum field is almost the same. To make curve `5', we compressed the baseline beam 4 times longitudinally and 4 times radially (by keeping the same emittance and population) and increased the plasma density 16 times (to keep $\sigma_r = c/\omega_p$). The resulting dimensionless wakefield is just 25\% higher.

The temporal growth of the wakefield for the baseline variant at the cross-section of the strongest modulation (at $z=2 \times 10^4 c/\omega_p = 4$\,m) is shown in Fig.\,\ref{f2-envelopes}(a). There is almost linear growth of the wakefield followed by fast drop. However, the reason for this field behavior is obscured by a complicated beam shape at the stage of developed self-modulation. To see the effect clearer, we simulate the field excitation by the train of rigid equidistant short bunched of the same radius and peak density:
\begin{align}
    \nonumber  n_b (r, z, t) &=  0.5\, n_{bm}\, e^{-r^2/2 \sigma_r^2} \left[  1 - \cos \bigl( 2 \omega_p (t-z/c)  \bigr)  \right], \\
    \nonumber   &\quad i \tau_0  < t-z/c < (i+1/2) \tau_0, \quad i=0, 1, \ldots ; \\
    \nonumber  n_b (r, z, t) &= 0, \qquad \text{otherwise},
\end{align}
where $\tau_0 = 2 \pi \omega_p^{-1}$.
The electric field and the field envelope for this test case are shown in Fig.\,\ref{f2-envelopes}(b). The behavior is qualitatively the same, but now the cause for the saturation is clearly seen in Fig.\,\ref{f2-envelopes}(c,d). The field stops growing when the wave shifts forward in time with respect to the drive bunches, so that the bunches mostly fall into the decelerating field. The key effect responsible for the field saturation is thus elongation of the wakefield period.

\begin{table}[b]
  \centering
  \caption{Simulated modes of wakefield excitation.}\label{t1}
  \begin{tabular}{cl}
    \hline \hline
    No. & Driver \\ \hline
    1 & Single bunch, $p^+$, variable charge, $\sigma_r=15\, c/\omega_p$ \\
    2 & Single bunch, $p^+$, variable charge, $\sigma_r=3\, c/\omega_p$ \\
    3 & Single bunch, $p^+$, variable charge, $\sigma_r=c/\omega_p$ \\
    4 & Train of 5 bunches, $p^+$, variable charge, $\sigma_r=c/\omega_p$ \\
    5 & Train of 5 bunches, $p^-$, variable charge, $\sigma_r=c/\omega_p$ \\
    6 & Train of 5 bunches, $p^+$, variable charge, $\sigma_r=0.3\, c/\omega_p$ \\
    7 & Train of 5 bunches, $p^-$, variable charge, $\sigma_r=0.3\, c/\omega_p$ \\
    8 & Infinite train, $p^+$, variable location, $\sigma_r=c/\omega_p$ \\
    9 & Infinite train, $p^-$, variable location, $\sigma_r=c/\omega_p$ \\ \hline
    A & Theory of one-dimensional wave \\
    B & Empirical approximation for the free wave\\
    C & Empirical approximation for the driven wave\\
    \hline \hline
  \end{tabular}
\end{table}
\begin{figure}[t]
\bc\includegraphics[width=207bp]{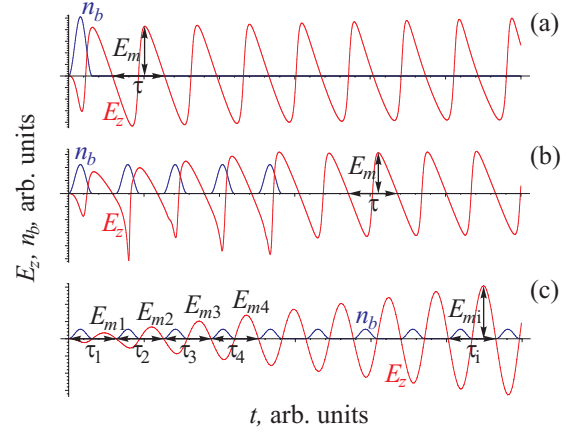} \ec
 \vspace*{-5mm}
\caption{(Color online) Location of the measured wave periods for the single bunch of variable charge (a), the train of 5 bunches (b), and the infinite bunch train (c). }\label{f3-details}
\end{figure}
For the two-dimensional (axisymmetric) plasma wave excited by charged particle bunches, the wakefield period depends not only on the wave amplitude, but also on the spacial structure of the wave which, in turn, depends on the driver shape. Moreover, the wave is not strictly periodical, and the wavelength changes as the distance from the driver increases. Unlike the one-dimensional plasma wave, there is no universal relationship between the maximum electric field and the wave period. To view the phenomenon broadly, we have simulated waves excited by various drivers (Table~\ref{t1}). We have measured either the period of a selected wave oscillation for drivers of certain shape and variable charge [Fig.\,\ref{f3-details}(a,b)], or periods of successive oscillations continuously driven by an infinite bunch train [Fig.\,\ref{f3-details}(c)].

\begin{figure}[t]
\bc\includegraphics[width=194bp]{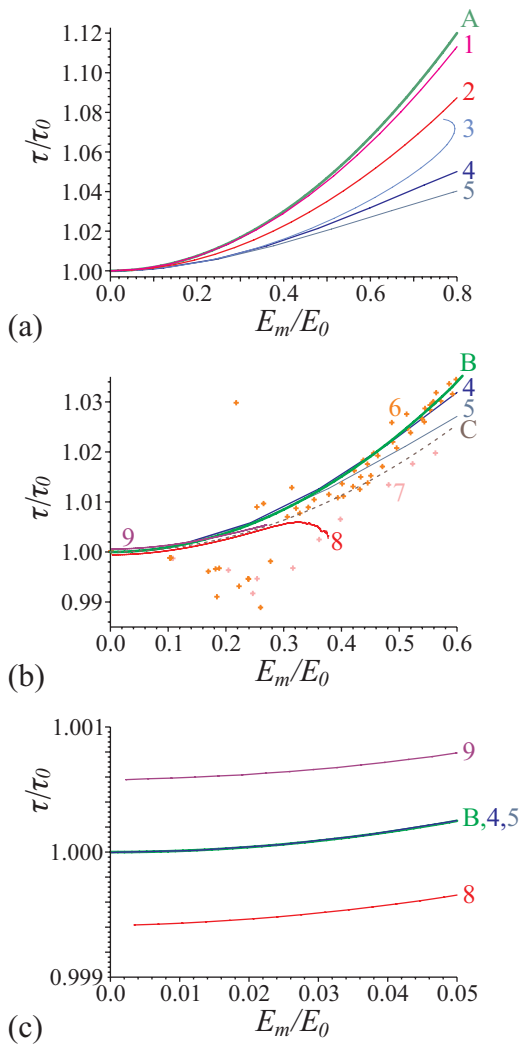} \ec
 \vspace*{-5mm}
\caption{(Color online) Dependence of the wakefield period on the wave amplitude for various drivers.}\label{f4-period}
\end{figure}
The results are shown in Fig.\,\ref{f4-period}. For all cases the wave period increases with the wakefield amplitude. The reason for that is the wave nonlinearity that mainly comes from relativistic corrections to the law of electron motion. The wider the driver, the stronger the wavelength dependence on the wakefield amplitude. In the limit of very wide beams (curve `1'), the result of one-dimensional analytical theory \cite{Akhiezer} is reproduced (line `A'):
\begin{equation}\label{e2}
    \tau \approx \tau_0 \left(1+\frac{3}{16}(E_m/E_0)^2\right).
\end{equation}
For narrow beams (cases `6' and `7') , the fields in the shown range are produced by very dense bunches and correspond to strongly nonlinear waves; the dependence $\tau (E_m)$ for these beams is not smooth and therefore is shown by dots in Fig.\,\ref{f4-period}(b). Curves `3', `4', and `5' in Fig.\,\ref{f4-period}(a) are for drivers of the same radius and show that the wave period does depend on the wave excitation method.

There are two other factors apart from the wave nonlinearity that have the effect on the wave period: the current compensation and the wave drive. The particle beam generates the plasma current opposite to the average current of the beam. The motion of plasma electrons associated with this current produces the Doppler shift of the plasma frequency which is independent on the wave amplitude. This effect is seen as shifted periods for the cases `8' and `9' at small wave amplitudes [Fig.\,\ref{f4-period}(c)]. For positively charged drivers, plasma electrons move forward, and the period is shorter; for negatively charged driver, the period is longer. If the beam is not too narrow ($\sigma_r \gtrsim c/\omega_p$), then the shift of the period is roughly the ratio of the length-average beam density to the plasma density; otherwise the beam density must be also averaged transversely over the area of the radius $c/\omega_p$. We emphasize that that effect of the plasma current is small only for the chosen small density of the drive beam and can dominate for denser bunches \cite{AIP396-75}.

The wave drive can change the distance between the field zeros, if the contributions of further bunches are phase shifted with respect to the already existing wave. As we see from Fig.\,\ref{f3-details}(c) or from the elementary theory of harmonic oscillators, this is not the case until the phase shift appears due to other effects. Consequently, the wave drive can only modify the law of variation of the wave period, but cannot produce the period change by itself. Difference of curves `8' and `9' from `4' and `5' in Fig.\,\ref{f4-period}(b) shows the quantitative effect of the wave drive.

As we see from the simulations, the dependence of the wakefield period on the  amplitude of the two-dimensional plasma wave can be approximated by a parabola with the coefficient $\alpha$ depending on driver width and shape:
\begin{equation}\label{e3}
    \tau \approx \tau_0 \left(1+\alpha (E_m/E_0)^2\right).
\end{equation}
For waves of the width $c/\omega_p$, we have $\alpha \approx 0.1$ (line `B').

We can estimate the maximum amplitude of the wave on the basis of the empirical formula (\ref{e3}). Denote $\Delta E$ the field increment due to one bunch. Then $N$ coherent bunches produce the field $E_z=N \Delta E$. The wave nonlinearity results in elongation of the wave period by
\begin{equation}\label{e4}
    \Delta \tau = \tau_0 \alpha N^2 (\Delta E/E_0)^2.
\end{equation}
The wave drive also makes a contribution to the wave period, which depends on the accumulated delay $T$ of the wave. A small increment $\Delta E e^{i\omega_p t}$ to the complex field $E_z e^{i\omega_p (t-T)}$ results in the small increment of the wave phase
\begin{equation}\label{e5}
    \delta \varphi \approx \frac{\Delta E \sin (\omega_p T)}{E_z + \Delta E \cos (\omega_p T)} \approx \frac{\omega_p T}{N},
\end{equation}
which means shortening of the wave period by $\omega_p^{-1} \delta \varphi$. The law of delay accumulation is thus
\begin{equation}\label{e6}
    T_N = \sum_{i=1}^N (\Delta \tau_i - T_i/i),
\end{equation}
or, in the limit of large $N$,
\begin{equation}\label{e7}
    \frac{dT}{dN} = \tau_0 \alpha (\Delta E/E_0)^2 N^2 - T/N.
\end{equation}
The solution to this equation is
\begin{equation}\label{e8}
    T = \tau_0 \alpha \left( \frac{\Delta E}{E_0}\right)^2 \frac{N^3}{4}.
\end{equation}
With no account of the wave drive, there would be the factor `3' in the denominator of (\ref{e8}) instead of `4'. In other words, the wavelength elongation for the driven wave is 25\% shorter that that for the free wave. Simulations confirm this observation as long as the wave is weakly nonlinear: lines `8' and `9' in Fig.\,\ref{f4-period}(b) follows the parabola `C' with $\alpha \approx 0.075$.

The wakefield stops growing if
\begin{equation}\label{e9}
   T = \beta \tau_0,
\end{equation}
where the factor $\beta$ depends on the wave shape. For the one-dimensional wave with no transverse structure, zero average deceleration of the bunches corresponds to $\beta \approx 1/4$. For essentially two-dimensional waves, $\beta \sim 1/2$ [Fig.\,\ref{f2-envelopes}(d)], since the bunches also interact with off-axis regions where the period mismatch is smaller.

The cubic dependence $T(N)$ is a fast growing function in a sense the transition between the cases $T \ll \tau_0$ and $T \sim \tau_0$ occurs quickly. Consequently, the field growth is linear [Fig.\,\ref{f2-envelopes}(b)] and formulae (\ref{e3})--(\ref{e8}) are valid up to the very instant of field saturation.

Equating expressions (\ref{e8}) and (\ref{e9}) yields the maximum number of coherent bunches
\begin{equation}\label{e10}
    N = \left(\frac{4\beta E_0^2}{\alpha \Delta E^2}\right)^{1/3}
\end{equation}
and the maximum field
 \begin{equation}\label{e11}
    E_\text{max} = \left(\frac{4\beta E_0^2 \Delta E}{\alpha}\right)^{1/3}.
 \end{equation}
Note the weak dependence (power 1/3) of these expressions on empirical parameters $\alpha$ and $\beta$ and on the contribution of a single bunch $\Delta E$.

For parameters of the test case ($\alpha=0.1$, $\beta=0.5$, $\Delta E = 0.003\,E_0$) the estimated maximum field $E_\text{max} \approx 0.4 \, E_0$ is quite close to simulations [Fig.\,\ref{f2-envelopes}(b)]. For the longitudinally compressed AWAKE-like beams (lines `4' and `5' in Fig.\,\ref{f1-growth}), no field increase $4^{1/3}$ times is observed since the number of bunches is less than required by (\ref{e10}).

This work is supported by The Ministry of education and science of Russia (projects 14.B37.21.0784, 14.B37.21.0750, and 8387) and by RFBR grants 11-01-00249 and 11-02-00563.

\end{document}